\documentclass[twocolumn,prl,floatfix,superscriptaddress,nobibnotes]{revtex4-1}

\usepackage{amsmath}
\usepackage{amssymb}
\usepackage{graphicx}
\usepackage{color}

\begin{document}

\title{Zero-field Topological Superconductivity in Ferromagnetic Hybrid Nanowires}

\author{S.~Vaitiek\.{e}nas}
\affiliation{Center for Quantum Devices, Niels Bohr Institute, University of Copenhagen, 2100 Copenhagen, Denmark}
\affiliation{Microsoft Quantum Lab--Copenhagen, 2100 Copenhagen, Denmark}

\author{Y.~Liu}
\affiliation{Center for Quantum Devices, Niels Bohr Institute, University of Copenhagen, 2100 Copenhagen, Denmark}
\affiliation{Microsoft Quantum Materials Lab--Copenhagen, 2800 Lyngby, Denmark}
\author{P.~Krogstrup}
\affiliation{Center for Quantum Devices, Niels Bohr Institute, University of Copenhagen, 2100 Copenhagen, Denmark}
\affiliation{Microsoft Quantum Materials Lab--Copenhagen, 2800 Lyngby, Denmark}
\author{C.~M.~Marcus}
\affiliation{Center for Quantum Devices, Niels Bohr Institute, University of Copenhagen, 2100 Copenhagen, Denmark}
\affiliation{Microsoft Quantum Lab--Copenhagen, 2100 Copenhagen, Denmark}

\date{\today}

\begin{abstract}
We report transport measurements and tunneling spectroscopy in hybrid nanowires with epitaxial layers of superconducting Al and the ferromagnetic insulator EuS, grown on semiconducting InAs nanowires. In devices where the Al and EuS covered facets overlap, we infer a remanent effective Zeeman field of order 1~T, and observe stable zero-bias conductance peaks in tunneling spectroscopy into the end of the nanowire, consistent with topological superconductivity at zero applied field.  Hysteretic features in critical current and tunneling spectra as a function of applied magnetic field support this picture. Nanowires with non-overlapping Al and EuS covered facets do not show comparable features. Topological superconductivity in zero applied field allows new device geometries and types of control.
\end{abstract}

\maketitle

Hybrid quantum materials allow for novel quantum phases that otherwise do not exist in nature \cite{Stern2010,Fu2008}. For example, a one-dimensional topological superconductor with Majorana zero modes at its ends can be realized by coupling a semiconductor nanowire to a superconductor in the presence of a strong Zeeman field \cite{Oreg2010,Lutchyn2010,Lutchyn2018}. However, the applied magnetic fields are detrimental to superconductivity, constraining device layout, components, materials, fabrication, and operation \cite{Karzig2017}. Early on, an alternate source of Zeeman coupling that circumvents these constraints---using a ferromagnetic insulator instead of an applied field---was proposed theoretically \cite{Sau2010}. 

Planar superconductor-ferromagnetic insulator (SC-FMI) hybrids exhibit spin-splitting of the superconducting density of states in zero applied field, reflecting ferromagnetic exchange coupling~\cite{Tedrow1986}, reminiscent of the spin-splitting generated by large in-plane magnetic fields~\cite{Meservey1970} (see Refs.~\cite{Izyumov2007, Bergeret2018, Heikkila2019}  for recent reviews). 
The ferromagnetic exchange coupling that results from spin-dependent scattering at the SC-FMI interface \cite{Tokuyasu1988} extends into the superconductor for a coherence length, while thinner superconductors become uniformly magnetized \cite{Bergeret2004}. The superconducting coherence length also averages over domain structure of the FMI film, modifying spin splitting depending on the domain size compared to coherence length \cite{Strambini2017}. Relaxation of exchange-induced spin splitting has been investigated using spin-polarized injection and detection in SC-FMI hybrids \cite{Wolf2014}.

While ferromagnetic exchange coupling in the Al occurs via spin-dependent scattering at the Al-EuS interface, in most contexts, exchange coupling can be thought of as arising from an effective Zeeman field within the superconductor, oriented along the magnetization direction of the EuS~\cite{Cottet2009}. This effect dominates over fringing field outside the FMI \cite{Liu2019_2}.
Proximity effect from the exchange-coupled superconductor into the spin-orbit coupled nanowire can induce a topological state in the hybrid system, depending on the arrangement of the interfaces \cite{Sau2010}. This mechanism contrasts, for example, recent work that uses spatially varying fringing fields from a nearby ferromagnet to synthesize a real-space spin-orbit field \cite{Desjardins2019}.
Recently, zero-bias peaks (ZBPs)---signature of Majorana modes---were observed in an applied magnetic field in Au nanowires with a superconductor proximitized surface state, using the same FMI, EuS, to tune the energy of the surface state closer to the Fermi energy~\cite{Manna2019}.
Materials studies of epitaxial EuS on InAs without Al ~\cite{Liu2019}, or with EuS and Al on adjacent (non-overlapping) wire facets \cite{Liu2019_2} showed weak ferromagnetic exchange coupling transferred to the InAs. Epitaxial growth of hybrid semiconductors with ferromagnetic-insulator and superconducting layers \cite{Liu2019_2} opens a new venue for topological superconductivity without the need for an applied magnetic field.

Wurtzite $[0001]$ InAs nanowires were grown by the vapor-liquid-solid (VLS) method using molecular beam epitaxy (MBE) \cite{Krogstrup2015}. Epitaxial EuS was then grown on two facets of the hexagonal wires, followed by Al, grown on two partly-overlapping or adjacent facets. Because Al requires low-temperature deposition, it was necessary to grow the EuS first \cite{Liu2019_2}. After placing wires on a back-gated Si substrate, contacts and gates were fabricated using electron beam lithography. Measurements used standard ac lock-in techniques in a dilution refrigerator with a three-axis vector magnet and base temperature of 20~mK (see Supplemental Material for fabrication and measurement details).

\begin{figure}
\includegraphics[width=\linewidth]{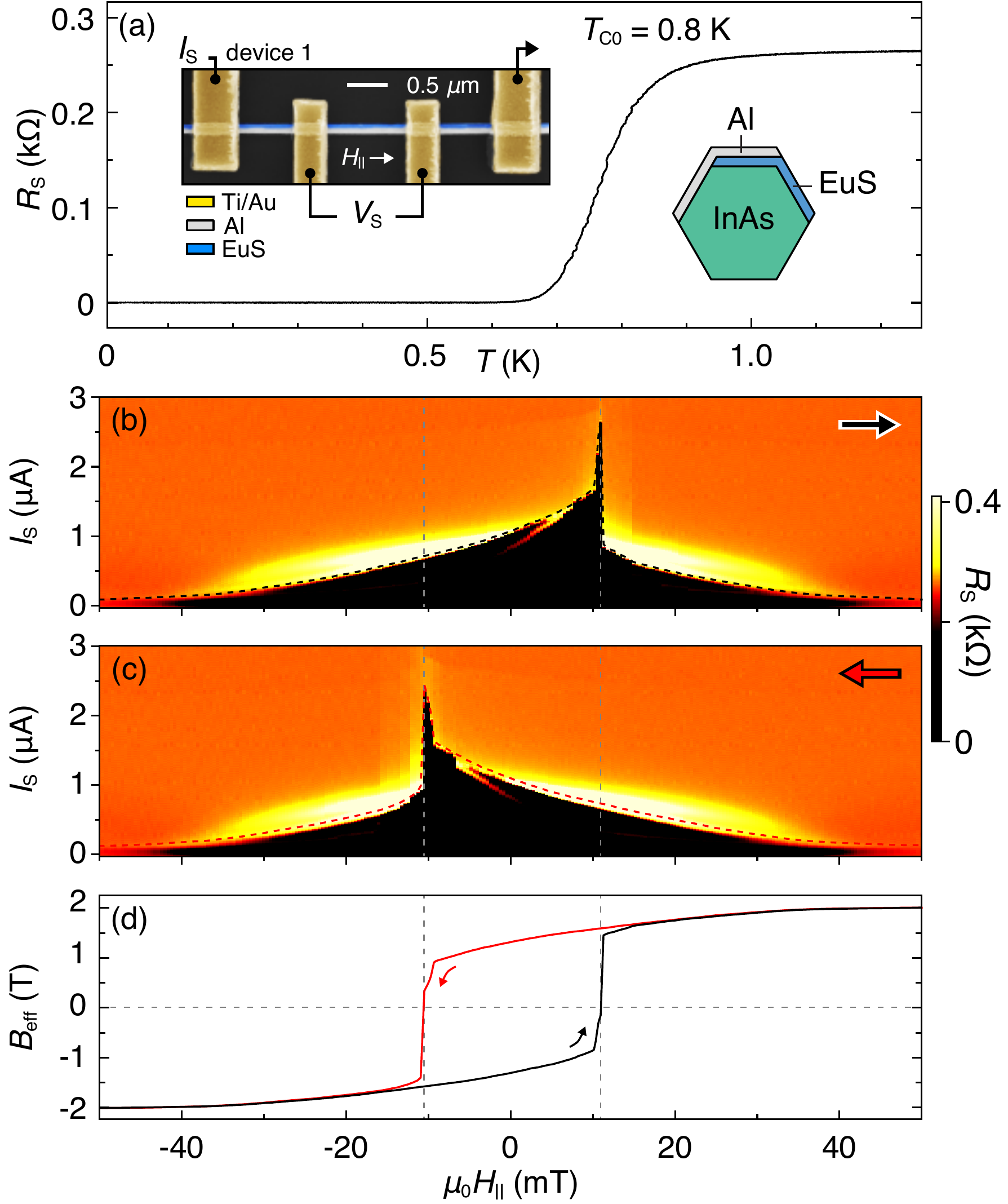}
\caption{\label{fig:1}
(a) Four-probe differential resistance of the Al/EuS shell, $R_{\rm S}$, measured for device 1 around zero bias as a function of temperature, $T$, shows a critical temperature $T_{\rm C0}\sim0.8$~K. Left Inset: Colorized micrograph of device 1 showing measurement set-up. Right inset: Schematic wire cross-section showing orientation of partly-overlapping EuS and Al shells.
(b,c) $R_{\rm S}$ as a function of applied magnetic field along wire axis, $H_\parallel$, and current bias, $I_{\rm S}$, sweeping $H_\parallel$ from (b) negative to positive and (c) positive to negative. (d) Effective Zeeman field, $B_{\rm eff}(H_\parallel)$ from pair-breaking analysis using $I_{\rm C}$ from (b) and (c), see Supplemental Material. Remanent effective field is $ B_{\rm eff}(H_{\parallel}=0)\sim 1.3$~T.
}
\end{figure}

\begin{figure*}[t!]
\includegraphics[width=\linewidth]{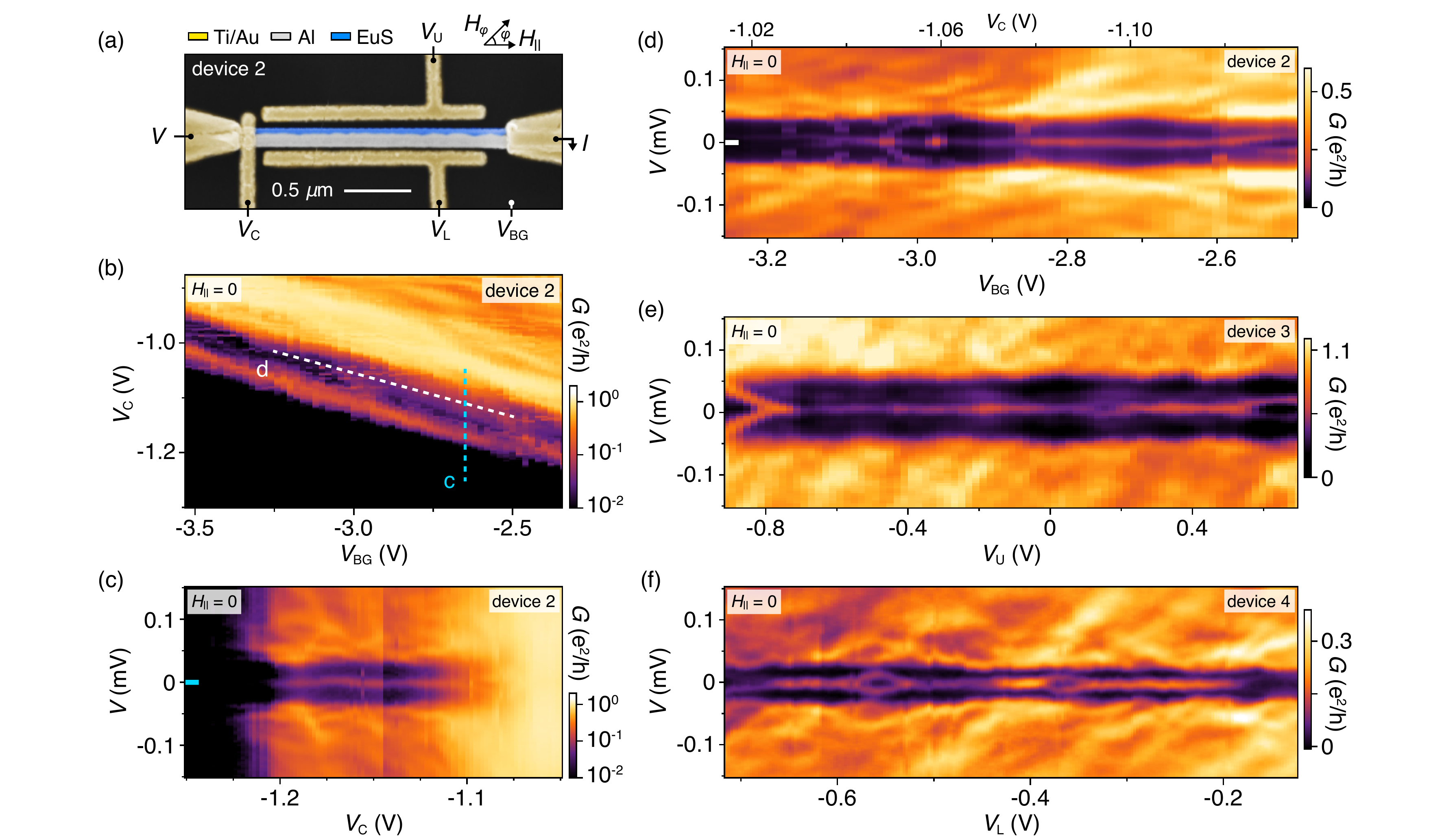}
\caption{\label{fig:2}
(a) Colorized micrograph of device 2 with measurement set-up. External magnetic field direction indicates the in-plane field angle $\varphi$. (b) Differential conductance, $G$, at zero bias as a function of barrier gate voltage, $V_{\rm C}$, and back-gate voltage, $V_{\rm BG}$, for device 2.
(c) $G$ as a function of source-drain bias voltage, $V$, and $V_{\rm C}$ from weak-tunneling to open regime, measured along the blue dashed line in (b). 
(d) $G$ as a function of $V$ and compensated $V_{\rm BG}$ measured along white dashed line in (b). Top axis shows compensation gate voltages.
(e, f) Similar to (d) for devices 3 and 4 as a function of uncompensated gate voltages $V_{\rm U}$ and $V_{\rm L}$, respectively. 
}
\end{figure*}

We present measurements on eight devices from two growth batches. Devices 1--6 have two-facet EuS and Al shells overlapping on one facet [Fig.~\ref{fig:1}(a), inset]; devices 7 and 8 have Al and EuS on adjacent non-overlapping pairs of facets [Fig.~\ref{fig:S1}(a), inset in Supplemental Material].
Devices 1 and 5 were used for four-terminal measurement of the shell [Fig.~\ref{fig:1}(a)]. Devices 2, 3, 4, and 6, used for tunneling spectroscopy [Fig.~\ref{fig:2}(a)], were lithographically equivalent, and showed similar behavior.
Roughly half of device batches showed subgap features in bias spectroscopy as reported here. Others showed no subgap features or a soft superconducting gap. Devices 7 and 8, with adjacent but non-overlapping Al and EuS shells, showed unsuppressed superconductivity in the Al shell, little dependence on field, and no subgap features (see Figs.~\ref{fig:S1} and \ref{fig:S2} in Supplemental Material). We draw attention to the observation that non-overlapping shells of EuS and Al do not produce the subgap features seen when the shells overlap. This is an important difference from an earlier proposal along these lines \cite{Sau2010}.

We begin by investigating the superconducting properties of the coupled Al/EuS shell. 
Differential resistance, $R_{\rm S} = {\rm d}V_{\rm S}/{\rm d}I_{\rm S}$, measured in a four-terminal configuration for device~1 as a function of temperature, $T$, yielded a zero-field critical temperature $T_{\rm C0} =0.8$~K [Fig.~\ref{fig:1}(a)], lower than the bulk value of 1.2~K.
These data were taken around zero dc current bias after ramping the external magnetic field applied along the wire axis to $\mu_0 H_\parallel = 150$~mT and then back to zero.
In addition to the reduced $T_{\rm C}$, the base-temperature critical current, $I_{\rm C}$, measured for device 1 displayed a characteristic evolution with $H_\parallel$ that depended on sweep direction [Figs.~\ref{fig:1}(b,c)].
Moving from negative to positive field, a finite $I_{\rm C}$ first appeared around $-50$~mT, increasing rapidly toward zero field. As the field passed through zero, $I_{\rm C}$ continued increasing, peaking around $+11$~mT, then rapidly decreasing and smoothly vanishing at around $+50$~mT [Fig.~\ref{fig:1}(b)].
The reverse evolution of $I_{\rm C}$ was observed when sweeping $H_\parallel$ from positive to negative, with the sharp peak at $\mu_0 H_\parallel = -11$~mT [Fig.~\ref{fig:1}(c)]. The reduced $T_{\rm C}$ and hysteretic behavior are consistent with an exchange coupling between the Al and the EuS, which becomes magnetized along the wire axis then switches direction hysteretically with a switching (coercive) field of $\sim\pm 11$~mT. We note that above the coercive field, forward and backward sweeps overlap, and do not show discontinuities or hysteresis in the onset of critical current at higher applied fields ($\sim50$~mT). This observation suggests a continuous decrease to zero of the superconducting order parameter, relevant for later discussion. 

As a control, similar measurements on a nanowire with Al and EuS on non-overlapping facets showed unsuppressed $T_{\rm C}=1.5$~K (enhanced compared to bulk due to thinness of Al \cite{Court2007}) and very little hysteresis (see Fig.~\ref{fig:S1} in Supplemental Material). The striking contrast between wires with overlapping versus adjacent non-overlapping shells of Al and EuS suggests that the relevant exchange coupling is directly between the EuS and the Al, and that the overlap is necessary for sizable exchange coupling.

From the dependence of $I_{\rm C}$ on $H_\parallel$ in Fig.~\ref{fig:1} and knowing $T_{\rm C0}$, we infer a related evolution of $T_{\rm C}$ with $H_\parallel$ using the relation $T_{\rm C} (H_\parallel) = T_{\rm C0} [I_{\rm C}(H_\parallel)/I_{\rm C0} ]^{2/3}$, where $I_{\rm C0}$ is the zero-field critical current~\cite{Bardeen1962} (see Supplemental Material).
The extracted $T_{\rm C}(H_\parallel)$ for device 1 (see Fig.~\ref{fig:S3} in Supplemental Material) yields $T_{\rm C0} = 0.8$~K, which increases to $\sim1.4$~K around the coercive magnetic fields, $\pm 11$~mT.
To estimate the effective Zeeman field generated by the interaction of Al with EuS, we associate the suppression of $T_{\rm C}$ (inferred from $I_{\rm C}$) with the magnetic interaction using a mean-field model, introducing a pair-breaking parameter, $\alpha(H_\parallel)$ \cite{Abrikosov1961, Tinkham1996}, and defining an effective Zeeman field proportional to the pair breaking parameter, $B_{\rm eff} =\alpha/\mu_{\rm B}$, appropriate for weak spin-orbit coupling in the Al \cite{Tinkham1996}. The Abrikosov-Gorkov expression for $T_{\rm C}(\alpha)$ yields the magnetization curve, $B_{\rm eff}(H_\parallel)$ shown in Fig.~\ref{fig:1}(d) (see Supplemental Material for details). This model is designed to give a rough estimate of effective field, and does not take into account, for instance, ballistic effects in clean Al \cite{Tokuyasu1988, Izyumov2007} or effects of spin-orbit coupling from the InAs or spin-flip scattering \cite{Bruno1973, Heikkila2019} from the rough Al-EuS interface, all of which may be important for a more detailed understanding of the system.  Within this picture, the estimated remanent effective field after returning to zero applied field is $\vert B_{\rm eff}(H_{\parallel}=0)\vert \sim 1.3$~T, consistent with previously measured values ~\cite{Hao1990,Xiong2011}. Zeeman fields of $\sim 1$~T were previously found sufficient to induce topological superconductivity in hybrid InAs wires without EuS \cite{Deng2016}. We note that there is considerable device-to-device variance in the critical field. For instance, the critical field for device 5 was 70~mT, as shown in Supplemental Fig.~\ref{fig:S4}, compared to 50~mT for device 1 in Fig.~\ref{fig:1}. 

Turning to bias spectroscopy, we measured differential conductance, $G = {\rm d}I/{\rm d}V$ across the gate-controlled tunnel barrier at the end of the hybrid nanowire into a normal contact as a function of source-drain bias, $V$ [Fig.~\ref{fig:2}(a)]. For weak tunneling, $G$ is proportional to the density of states at the end of the wire convolved with temperature \cite{Stanescu2011, Mourik2012, Deng2016, Nichele2017, Zhang2018}. 
The tunnel barrier was tuned using gate voltage $V_{\rm C}$. Carrier density and spatial distribution of carriers in the nanowire were tuned with a combination of upper ($V_{\rm U}$), lower ($V_{\rm L}$), and back-gate ($V_{\rm BG}$) voltages. Because the back-gate extends under the barrier, changes in $V_{\rm BG}$ had to be compensated by small changes in $V_{\rm C}$ to maintain tunneling rate and the occupancy of any resonances in the tunnel barrier, as illustrated for device 2 in Fig.~\ref{fig:2}(b). Compensation of $V_{\rm U}$ and $V_{\rm L}$ was not necessary.
Conductance in device 2 measured along the dashed blue line in Fig.~\ref{fig:2}(b) as a function of $V_{\rm C}$ ranges from weak tunneling, $G\ll e^2/h$, to the open regime, $G>e^2/h$, see Fig.~\ref{fig:2}(c).
Similar sweeps at different $V_{\rm BG}$ are shown in the Supplemental Fig.~\ref{fig:S5}. 
Bias spectra show a characteristic superconducting gap $\Delta \sim 50 \, \mu$eV with a single peak at zero bias extending from $V_{\rm C} = -1.2$~V to $-1.1$~V. The induced gap is considerably smaller than the corresponding gap, $\Delta\sim 230\, \mu$eV, measured in wires with non-overlapping Al and EuS shells (see Fig.~\ref{fig:S2} in Supplemental Material). For increased tunneling, the zero-bias peak evolved into a zero-bias dip that saturates near $G\sim 2e^2/h$ after splitting (see Fig.~\ref{fig:S5} in Supplemental Material), consistent with topological superconductivity~\cite{Vuik2018}. As shown in Supplemental Fig.~\ref{fig:S5}, the peak-to-dip crossover occurs for conductance values ranging from $0.2 \, e^2/h$ to above $e^2/h$ over a range of back-gate voltages. In a topological wire, this variation may result from disorder or multiple channels in the junction. Trivial subgap states at zero energy typically do not show such a crossover~\cite{Vuik2018}.

Bias spectra measured along a compensated cut of $V_{\rm BG}$ [white dashed line in Fig.~\ref{fig:2}(b)] at zero applied field are shown in Fig.~\ref{fig:2}(d). Spectra show subgap Andreev states that coalesce to zero bias at $V_{\rm BG} = -3.1$~V and remain at zero bias until $V_{\rm BG} = -2.6$~V before splitting again. Several sharp resonances---features descending from the gap and crossing zero energy---are visible in the sweep. These resonances depend only on $V_{\rm C}$ [horizontal peaks in Fig.~\ref{fig:2}(b)] indicating that they arise from states in or near the tunnel barrier.
Importantly, as these end-state resonances cross or anticross at zero bias, the extended ZBP itself does not split.
More examples of the non-splitting of the ZBP across resonances are shown in Supplemental Fig.~\ref{fig:S6}.
Non-splitting of a ZBP through an end-state anticrossing is evidence of topological superconductivity and provides a bound on overlap between Majorana modes~\cite{Clarke2017, Prada2017}, as investigated previously~\cite{Deng2018}.

\begin{figure}[t!]
\includegraphics[width=\linewidth]{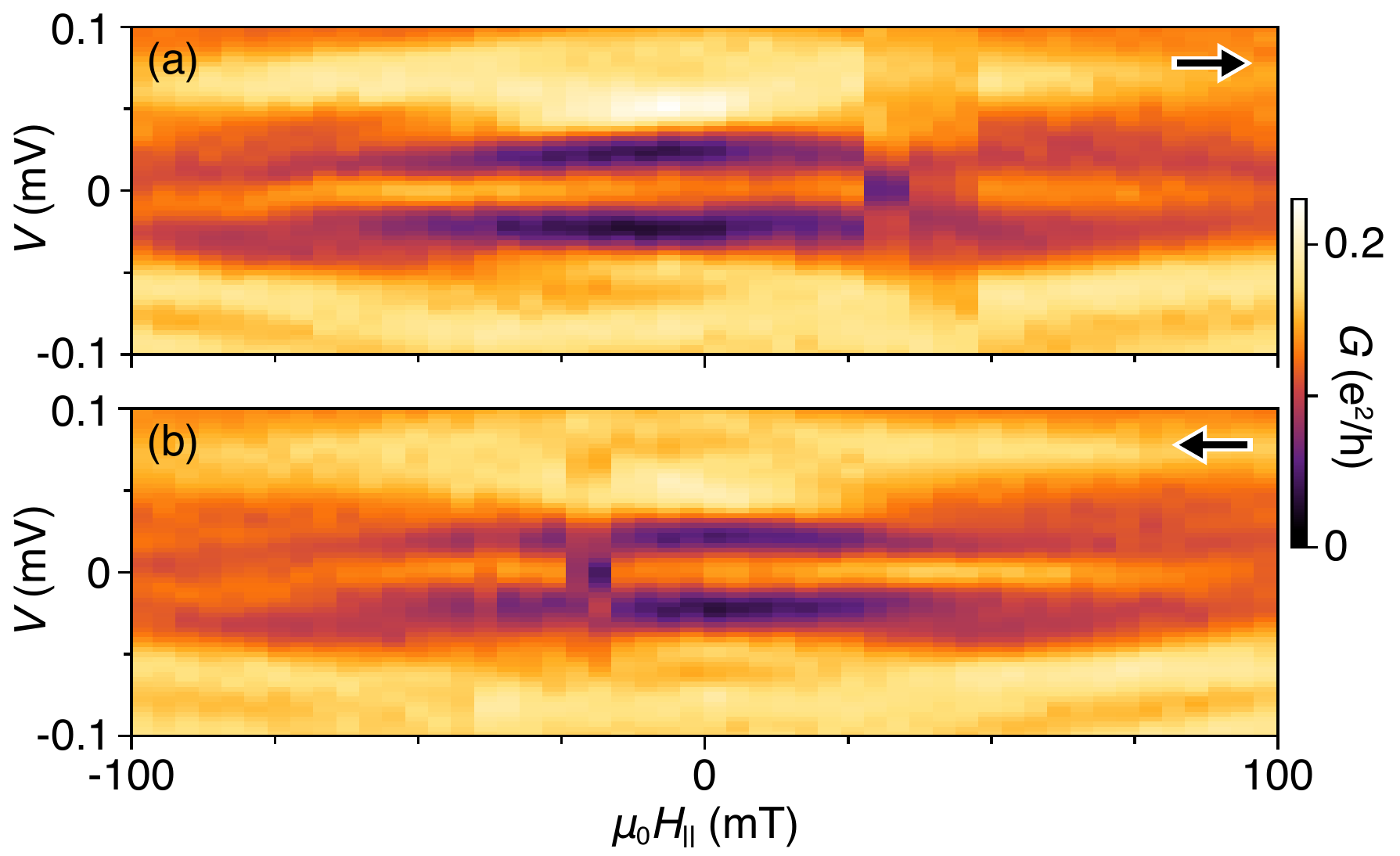}
\caption{\label{fig:3}
(a) Differential conductance, $G$, as a function of source drain voltage bias, $V$, and applied axial magnetic field, $H_\parallel$, for device~4 near the zero-bias peak splitting at $V_{\rm L} = -0.6$~V in Fig.~\ref{fig:2}(f). Sweep direction indicated by arrow. Zero-bias peak splits around the positive coercive field, $+25$~mT, where $|B_{\rm eff}|$ is minimized. 
(b) Same as (a) with sweep direction from positive to negative. Zero-bias peak splits at negative coercive field, $-25$~mT. 
}
\end{figure}

Figures~\ref{fig:2}(e,f) show bias spectra for other devices, measured as a function of uncompensated side-gate voltages $V_{\rm U}$ (device 3) and $V_{\rm L}$ (device 4). Both devices display stable ZBPs over a range of chemical potential of $\sim 0.4$~meV, estimated from the lever arms of the subgap states before merging at zero energy \cite{Deng2018}. Line-cuts of data in Figs.~\ref{fig:2}(c--f) as well as stability of the ZBP with $V_{\rm U}$ and $V_{\rm L}$ are shown in Supplemental Figs.~\ref{fig:S5}--\ref{fig:S7}.

Occasional splittings of the ZBP were seen in most of the sweeps, suggesting that the effective field does not greatly exceed the field needed to drive subgap states to zero energy. The gate dependence of splitting presumably reflects merging of the subgap states that have barely made it to zero, and are sensitive to the smooth disorder potential which affects both Majorana separation and overlap of residual spin character if not widely separated \cite{Vuik2018}. Gate-dependent splitting and rejoining argues for a discrete subgap state rather than a so-called class {\it D} peak, which is composed of many overlapping subgap states in disordered superconductors \cite{Brouwer2011, Bagrets2012, Sau2013} and so is not expected to readily split and remerge at zero.

Tuning to gate configurations near a splitting rendered the subgap spectrum particularly sensitive to $B_{\rm eff}$. This is illustrated in Fig.~\ref{fig:3}, where sweep-direction-dependent splittings are seen near the coercive fields for this device, $\pm 25$~mT, where $B_{\rm eff}$, averaged over the coherence length, is reduced when the magnetization reverses. The discontinuous jump in magnetization at the coercive field [see Fig.~\ref{fig:1}(d)] means that the average effective field is always large, even during a field reversal. The splitting seen in Fig. 3 only happens at gate voltages where the slightly reduced average effective field near the reversal point is insufficient to keep the peak at zero bias. More commonly, ZBPs are robust to a decrease in effective field, and so remains intact through the discontinuous reversal of magnetization.

\begin{figure}[t!]
\includegraphics[width=\linewidth]{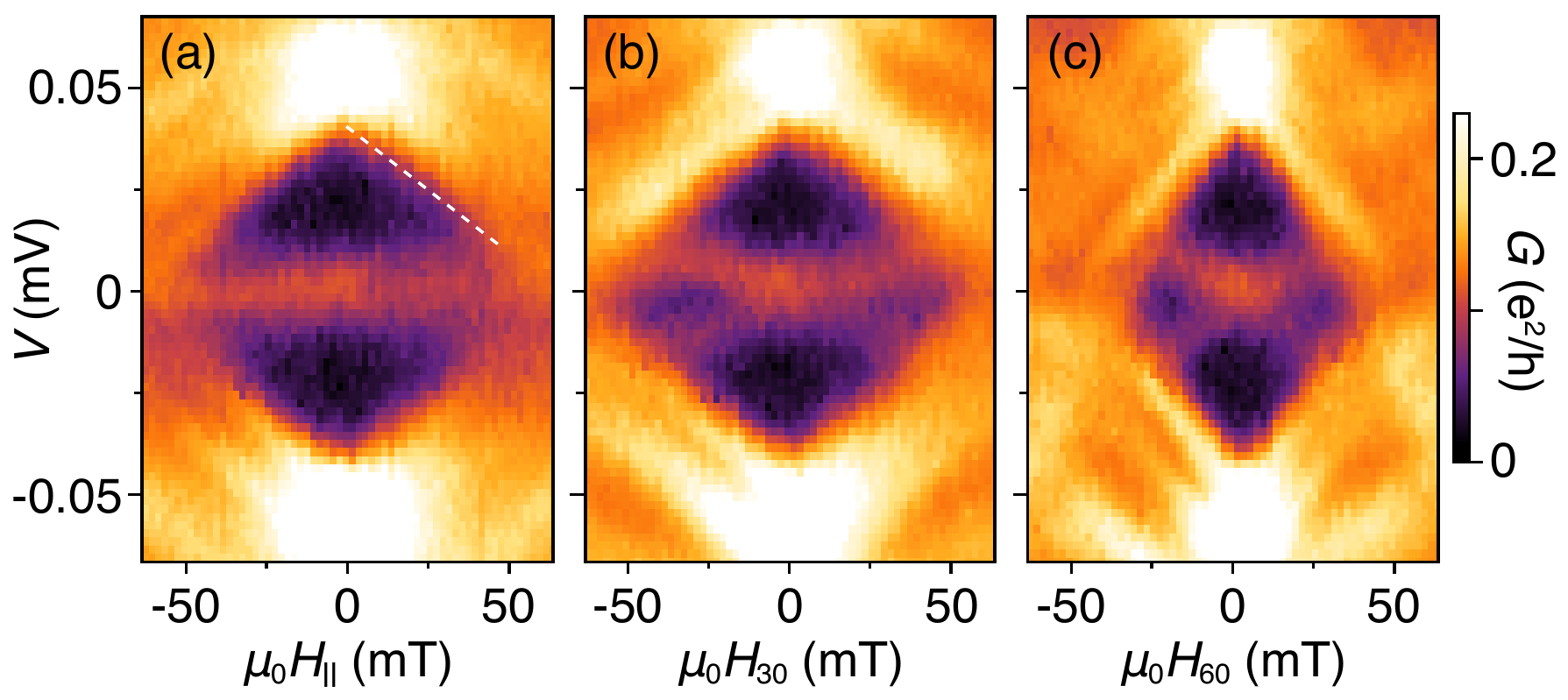}
\caption{\label{fig:4}
(a) Differential conductance, $G$, as a function of voltage bias, $V$, and axial magnetic field, $H_\parallel$, for device~2. Bias spectra show persistent zero-bias peak until the gap collapses at $\sim 50$~mT as the gap collapses. Note symmetry in field. The white dashed line corresponds to an effective $g$ factor of $\sim 20$.
(b) Similar to (a) with field $H_{30}$ in the wire plane, 30 degrees off axis. The zero-bias peak is split for $\mu_{0}\vert H_{30} \vert > 20$~mT.
(c) Similar to (b) with field $H_{60}$ applied 60 degrees off axis. The zero-bias peak is split for $\mu_{0}\vert H_{60} \vert > 10$~mT.
All data taken at a gate configuration corresponding to $V_{\rm U} = -0.6$~V in Supplemental Fig.~\ref{fig:S6}(i) with field swept from negative to positive.}
\end{figure}

An applied axial field rapidly decreases the superconducting gap but typically does not affect the ZBP, as illustrated in Fig.~\ref{fig:4}(a).  However, the same ZBP can be split by an applied off-axis field, as shown in the other panels of Fig.~\ref{fig:4}. For in-plane angles $\varphi = 30(60)$~degrees [see Fig.~\ref{fig:2}(a) for orientation] the ZBP is clearly split beyond an angle-dependent field, $\mu_{0}|H_{30(60)}|>20(10)$~mT. The large effective $g$ factor, exceeding 20, measured from the slopes of the gap edge [see dashed line in Fig.~\ref{fig:4}(a)] is presumably dominated by the suppression of superconductivity by the EuS magnetization, which is field dependent around zero applied field [see Fig.~\ref{fig:1}(d)]. Depending on tuning, the ZBP can remain stable for all angles of a sizable applied field (30 mT) as illustrated in Supplemental Fig.~\ref{fig:S8}. 

Splitting of ZBPs from an off-axis field suggests that the spin-orbit field is oriented transverse to the wire axis ~\cite{Nijholt2016}, as expected for Rashba-type spin-orbit coupling.  Measurements over a range of $\varphi$ show that the coercive field increases as $1/\cos ({\varphi})$ (see Fig.~\ref{fig:S9} in Supplemental Material), indicating that the EuS magnetization is predominantly along the wire axis, and that the component of applied field along the wire controls axial magnetization, as expected for this highly anisotropic geometry with easy magnetization axis along the wire.  Upon returning to zero field, the ZBP is recovered for all $\varphi$. This allows an off-axis fields to be used to magnetize the EuS, which is important for initializing non-collinear wires using a single-axis magnet.

Despite strong hysteretic effects in the EuS/Al shell (see Fig.~\ref{fig:1}), the field-dependent quasiparticle gaps in Fig.~\ref{fig:4} appear symmetric around $H_\varphi = 0$ for all field angles. This does not reflect a lack of visibility of subgap hysteresis, but reflects a real difference between the field dependence of the shell and the subgap states, consistently observed in all samples. We tentatively attribute this symmetry to spin mixing due to spin-orbit coupling in the wire~\cite{Tedrow1971,Bruno1973}.

Finally, we note that the subgap states that are already split at zero applied field can move either toward or away from zero energy with  applied field, as shown in Supplemental Fig.~\ref{fig:S10}. The slopes of these subgap states are typically much less than the slope of the gap closure, consistent with picture of rapid suppression of superconductivity due to the field-dependent EuS magnetization.

In summary, we have measured transport and bias spectroscopy in hybrid InAs-core nanowires with overlapping epitaxial shells of superconductor (Al) and ferromagnetic insulator (EuS). An inferred remanent Zeeman field of $\sim 1.3$~T along the wire axis in the Al/EuS shell gives rise to robust zero-bias peaks. Dependence on the overlap of the Al and EuS shells, characteristic hysteresis in applied field, stability through resonances, peak-to-dip crossover at large conductance, and stability of zero bias peaks over extended range of gate voltages, together argue for an interpretation of topological superconductivity at zero applied field, driven by combined proximity effects of magnetization from EuS and superconductivity from epitaxial Al.  The significance of these results to information technology include the ability to control effective magnetic fields on submicron length scales, an effective field that always follows the axis of the nanowire allowing  non-collinear topological wires, the ability to operate on-chip superconducting components at zero field, and elimination of the need for precise alignment of external fields.

We thank Z.~Cui, I.~P.~Zhang, and K.~A.~Moler for EuS magnetization studies, C.~S\o rensen for contributions to materials growth, S.~Upadhyay for nanofabrication, and K.~Flensberg for valuable discussions. Research was supported by Microsoft, the Danish National Research Foundation, and the European Commission.


\bibliography{bibfile}
\bibliographystyle{apsrev4-1}

\onecolumngrid
\clearpage
\onecolumngrid
\setcounter{figure}{0}
\setcounter{equation}{0}
\section{\large{S\MakeLowercase{upplemental} M\MakeLowercase{aterial}
}}
\renewcommand\thefigure{S\arabic{figure}}
\renewcommand{\tablename}{Table.~S}
\renewcommand{\thetable}{\arabic{table}}
\twocolumngrid



\section{Methods}

\subsection{Nanowire growth} InAs nanowire growth was catalyzed by Au on InAs (111)B substrate at $447~^\circ$C via the vapor-liquid-solid method using molecular beam epitaxy~\cite{Krogstrup2015}. Hexagonal InAs wires with wurtzite crystal structure were grown to a length of ${\sim}10~\mu$m along $[0001]$ with a planar growth rate of $0.5~\mu$m/hr and the As$_{2}$/In flux ratio of 28. The wire diameter was then thickened to $\sim100$~nm by further InAs radial growth at $350~^\circ$C with As$_{2}$/In flux ratio of 22.5. Next, EuS shell (8~nm for devices 1--4; 5~nm for devices 5 and 6) was grown \textit{in~situ} on two of the six wire facets at $180~^\circ$C using electron beam evaporation with an average growth rate of 0.02~nm/s~\cite{Liu2019,Liu2019_2}. For devices 5 and 6 a 10~nm AlOx protecting layer was grown on EuS. Subsequently, the Al shell (6~nm for all devices) was grown at $-30~^\circ$C with a deposition rate of 0.04~nm/s after rotating the wires by $60^\circ$ (devices 1--4) or $120^\circ$ (devices 5 and 6) with respect to the metal source resulting in partly-overlapping [Fig.~\ref{fig:1}(a)] or non-overlapping [Fig.~\ref{fig:S1}(a)] EuS and Al facets.

\subsection{Sample preparation} Devices were fabricated using standard electron beam lithography on a degenerately n-doped Si chip with 200~nm SiOx capping.
Individual hybrid wires were transferred from the growth substrate onto the fabrication chip using a manipulator station with a tungsten needle.
To contact the Al shell in devices 1, 5 and 7 we used A6 PMMA resist, milled the wires with AR-ion plasma for 9~min at 25~W and then deposited normal Ti/Au (5/150~nm) leads.
We found that the quality of Al etching improved if the wires were coated with a thin layer of AR 300-80 adhesion promoter.
For selective Al wet etch in devices 2, 3, 6 and 8 we used EL6 resist and MF-321 photoresist developer with 30~s etching time at room temperature;
in device 4 we used CSAR 62 (AR-P 6200) 13\% resist and IPA:TMAH 0.2~N developer 19:1 solution with 4~min room temperature etching time.
Normal metal Ti/Au (5/150~nm) contacts to InAs in devices 2, 3, 4, 6 and 8 were metalized using A6 PMMA resist after 7~min Ar-ion plasma milling at 15~W.
A thin layer (8~nm) of HfO$_2$ was grown using atomic layer deposition on devices 2, 3, 4 and 6.
Ti/Au (5/150~nm) gate electrodes in devices 2, 3, 4, 6 and 8 were deposited using A6 PMMA resist.

\subsection{Measurements}

Transport measurements were performed by standard ac lock-in techniques at 129~Hz in a XLD-400 Bluefors dilution refrigerator with a base temperature of 20~mK, equipped with a three axis (1, 1, 6)~T vector magnet. The dc lines that were used to measure and gate the devices were equipped with RF and RC filters adding an additional line resistance of $R_{\rm line} = 6.7$~k$\Omega$. Four-terminal differential resistance measurements in devices 1 and 5 were carried out using an ac current excitation of $100$~nA. Two-terminal differential conductance measurements for bias spectroscopy in devices 2, 3, 4 and 6 were performed with $5$~$\mu$V ac-voltage excitation.

\subsection{Critical temperature measurements}

The superconducting transition temperature for device 1 (with partly-overlapping EuS and Al shells), $T_{\rm C0} = (0.8 \pm 0.1)$~K, was deduced by Gaussian fit to the numerically differentiated data shown in Fig.~\ref{fig:1}(a); the uncertainty is given by the half width at half maximum of this rather broad transition peak. To gather statistics we carried out the same measurements on two additional wires lithographically equivalent to device 1 (not discussed in the main text) giving mean $T_{\rm C0} = (0.90 \pm 0.07)$~K, where the uncertainty is the standard deviation of the mid-point transition temperatures among these wires.

Similarly, the data in Supplemental Fig.~\ref{fig:S1}(a) yielded $T_{\rm C0} = (1.53 \pm 0.05)$ for device 5 (with non-overlapping EuS and Al shells), which together with two additional, lithographically equivalent devices (not discussed in the maing text) gave mean $T_{\rm C0} = (1.50 \pm 0.02)$~K.

\subsection{Hysteresis loop}

The mean-field theory of Abrikosov and Gorkov~\cite{Abrikosov1961} was originally developed for suppression of superconductivity by magnetic impurities. It was later shown to be valid for any case where $T_{\rm C}$ is reduced by a pair-breaking perturbation, for example an external magnetic field, see the discussion in Section 10.2 of Ref.~\cite{Tinkham1996}.
Application of Abrikosov-Gorkov theory to the specific problem of an interface between a superconductor and a ferromagnetic insulator expressed the effective scattering of electrons at the interface in terms of an effective Zeeman field inside the superconductor \cite{Tokuyasu1988}. The pair-breaking parameter $\alpha$ was then treated as resulting from this effective Zeeman field instead of a more complex scattering problem. For superconductors thin compared to their coherence length the effective field is uniform in the superconductor.
The transition temperature $T_{\rm C}$ depends on the pair-breaking parameter $\alpha$ as

\begin{equation}\label{eq:digamma}
    \ln\left(\frac{T_{\rm C}(\alpha)}{T_{\rm C}(\alpha = 0)}\right) 
    = \Psi\left(\frac{1}{2}\right) 
    - \Psi\left( \frac{1}{2} + \frac{\alpha}{2\pi k_{\rm B} { }T_{\rm C}(\alpha)} \right),
\end{equation}

\noindent where $\Psi$ is the digamma function and $T_{\rm C}(\alpha = 0)$ is the unperturbed critical temperature. We take  $T_{\rm C}(\alpha = 0) = 1.5$~K based on measurements on wires with non-overlapping shells of similar Al thickness, see Supplemental Figs.~\ref{fig:S1} and \ref{fig:S3}.
Assuming weak spin-orbit coupling in the Al, a thin Al shell, and purely axial applied field, we take a Zeeman form for the pair-breaking parameter, $\alpha = \mu_{\rm B} B_{\rm eff}$~\cite{Tinkham1996} with an effective Zeeman field, $B_{\rm eff} = \mu_0(M + H_\parallel)$, where $\mu_0$ is vacuum permeability, $M$ is magnetization, and $H_\parallel$ is the applied magnetic field. 
As described in the main text, we infer $T_{\rm C}(H_\parallel)$ from $I_{\rm C}(H_\parallel)$ using the relation \cite{Bardeen1962}

\begin{equation}\label{eq:tc}
    T_{\rm C} (H_\parallel) = T_{\rm C0} \left(\frac{I_{\rm C}(H_\parallel)}{I_{\rm C0}} \right)^{2/3},
\end{equation}

\noindent with $H_{\parallel}=0$ values $T_{\rm C0} = 0.8$~K and $I_{\rm C0} = 1.05$~$\mu$A from Figs.~\ref{fig:1}(a, b) as inputs.

Equation~\eqref{eq:digamma} is then evaluated numerically yielding $B_{\rm eff}(H_{\parallel})$ shown in Fig.~\ref{fig:1}(d), where the sign of $B_{\rm eff}(H_{\parallel)}$ is given by the slope of $\alpha$ (see Supplemental Fig.~\ref{fig:S3}). Note the absence of fitting parameters in this analysis.

\subsection{Correcting for line resistance}

The measured two-terminal conductance include the additional in-line filter resistance.
In the line-cuts shown in Supplemental Fig.~\ref{fig:S5} and \ref{fig:S7} we correct for the voltage drop over the filters by, first, numerically integrating the conductance-voltage ($G$--$V$) curves giving current-voltage ($I$--$V$) curves;
then, subtracting the product of the calculated current and line resistance from the set voltage ($V-IR_{\rm line}$);
finally, numerically differentiating the data again yielding corrected conductance-voltage ($\vphantom{\frac{1}{1}^1}\widetilde{G}$--$V$) curves.

\onecolumngrid
\clearpage

\section{Supplemental Figures}

\begin{figure}[h!]
\includegraphics[width=\linewidth]{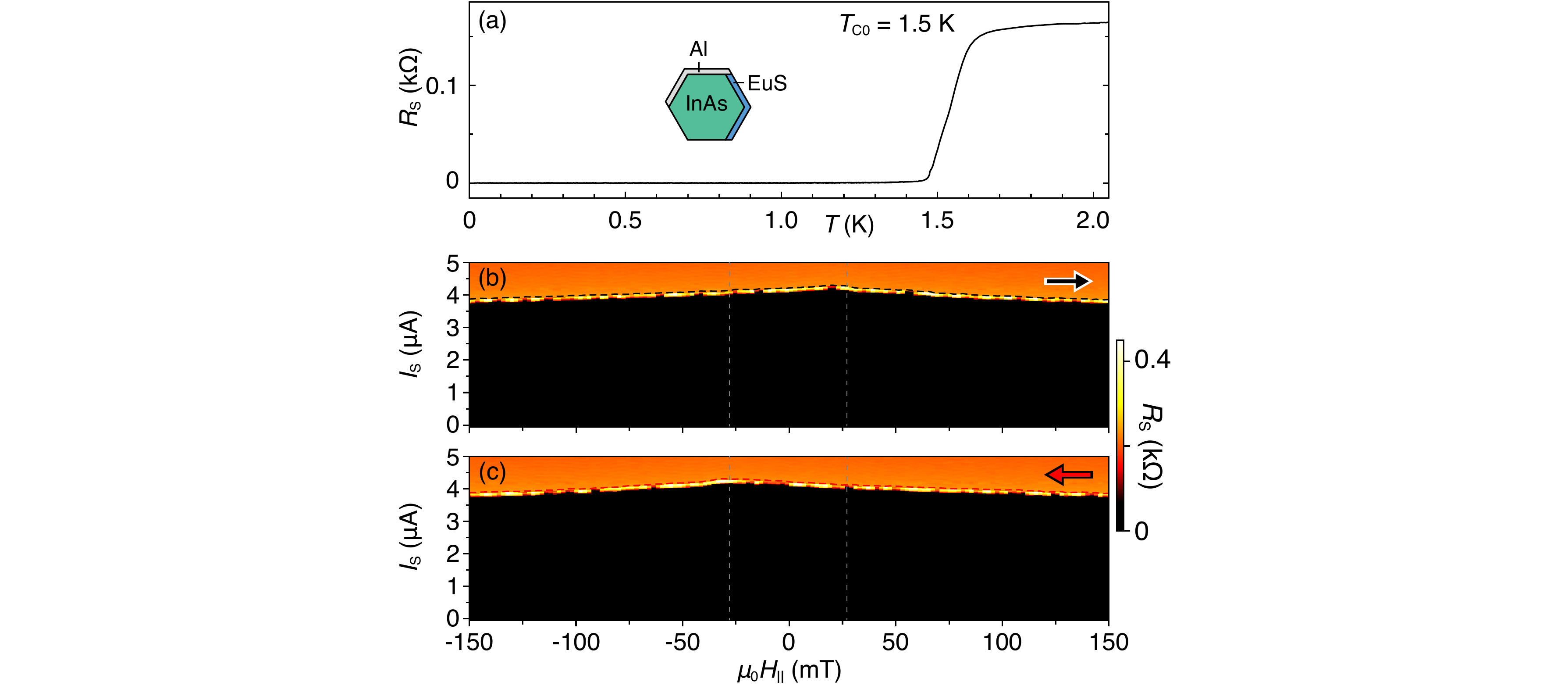}
\caption{\label{fig:S1}
(a) Four-probe differential resistance of the Al shell adjacent to EuS, $R_{\rm S}$, measured for device 7 around zero bias as a function of temperature, $T$, shows a critical temperature $T_{\rm C0}\sim1.5$~K.
Inset: Schematic wire cross section showing orientation of Al and EuS shells on adjacent pairs of facets.
(b,c) $R_{\rm S}$ as a function of applied magnetic field along wire axis, $H_\parallel$, and current bias, $I_{\rm S}$, sweeping $H_\parallel$ from (b) negative to positive and (c) positive to negative.
}
\end{figure}

\begin{figure}[h!]
\includegraphics[width=\linewidth]{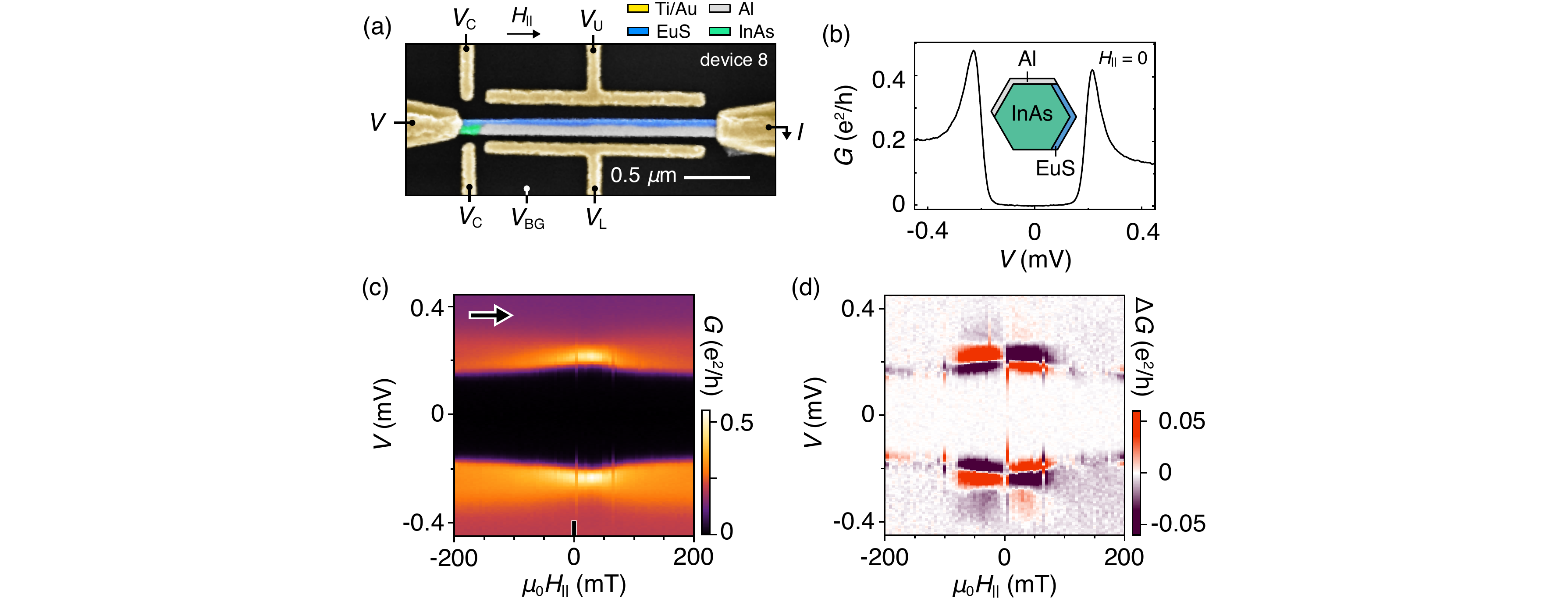}
\caption{\label{fig:S2}
(a) Colorized micrograph of device 8 with the measurement set-up. External axial magnetic field, $H_\parallel$, is indicated by an arrow. 
(b) Differential conductance, $G$, as a function of source-drain bias voltage, $V$, at $H_\parallel = 0$ shows a hard induced superconducting gap around $\Delta = 230~\mu$eV.
Inset: Schematic wire cross section showing orientation of Al and EuS shells on adjacent pairs of facets.
(c) Evolution of the tunneling spectrum with $H_\parallel$ swept from negative to positive.
(d) Difference of two conductance maps taken with $H_\parallel$ swept to opposite directions illustrates a weak hysteresis. The data were taken at $V_{\rm C} = -6.65$~V, $V_{\rm U} = -7$~V and $V_{\rm L} = 1.65$~V.
}
\end{figure}

\begin{figure}[p!]
\includegraphics[width=\linewidth]{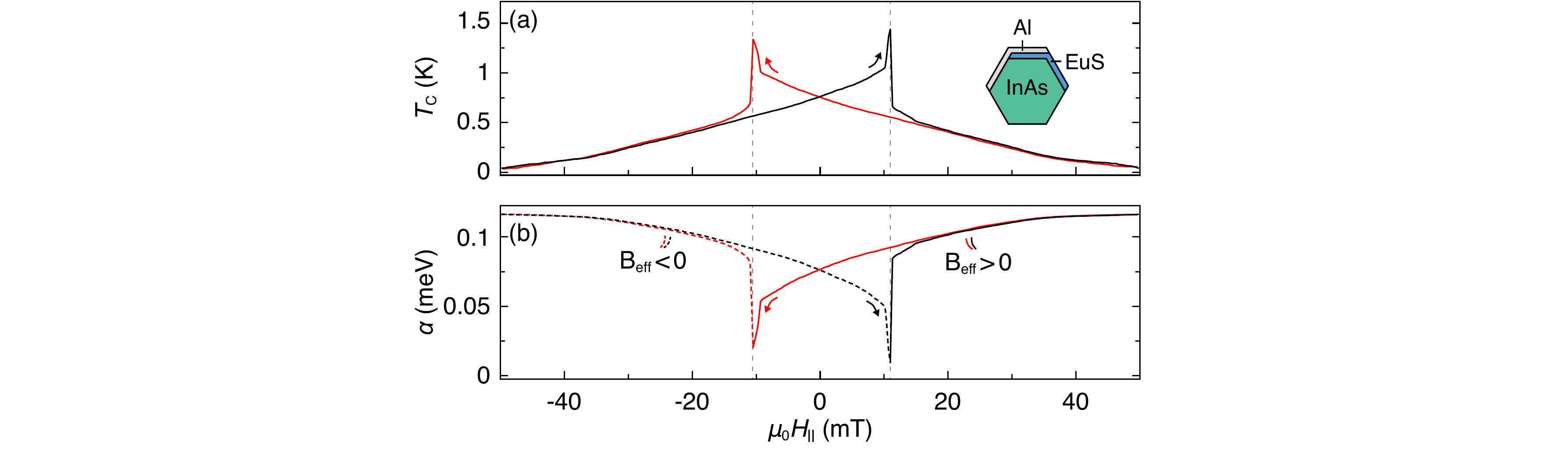}
\caption{\label{fig:S3}
(a) Critical temperature, $T_{\rm C}$, for device 1 as a function of applied axial magnetic field, $H_\parallel$, deduced from the critical current fits shown in Figs.~\ref{fig:1}(b,c) using Eq.~\eqref{eq:tc}. (b) Pair-breaking parameter, $\alpha$, estimated using Eq.~\eqref{eq:digamma} as described in Methods.
}
\end{figure}

\begin{figure}[p!]
\includegraphics[width=\linewidth]{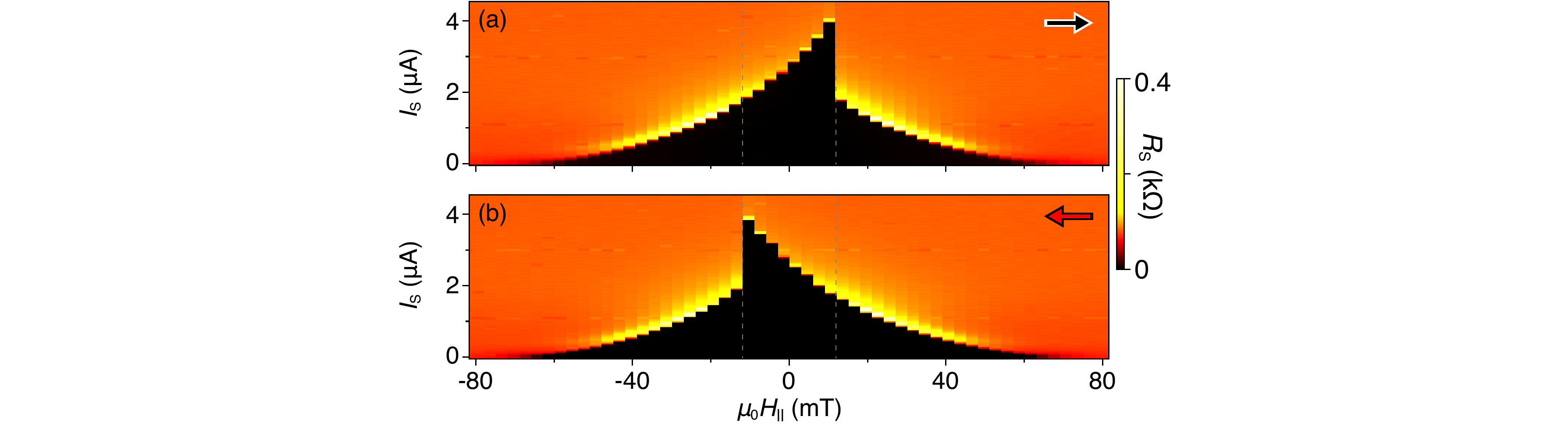}
\caption{\label{fig:S4}
Four-probe differential resistance of the Al/EuS shell, $R_{\rm S}$, measured for device 5 [lithographically equivalent to device 1 shown in Fig.~\ref{fig:1}(a)] as a function of applied magnetic field along wire axis, $H_\parallel$, and current bias, $I_{\rm S}$, sweeping $H_\parallel$ from (a) negative to positive and (b) positive to negative,
displays hysteretic behaviour and a critical current vanishing at around $\pm 70$~mT.}
\end{figure}

\begin{figure}[p!]
\includegraphics[width=\linewidth]{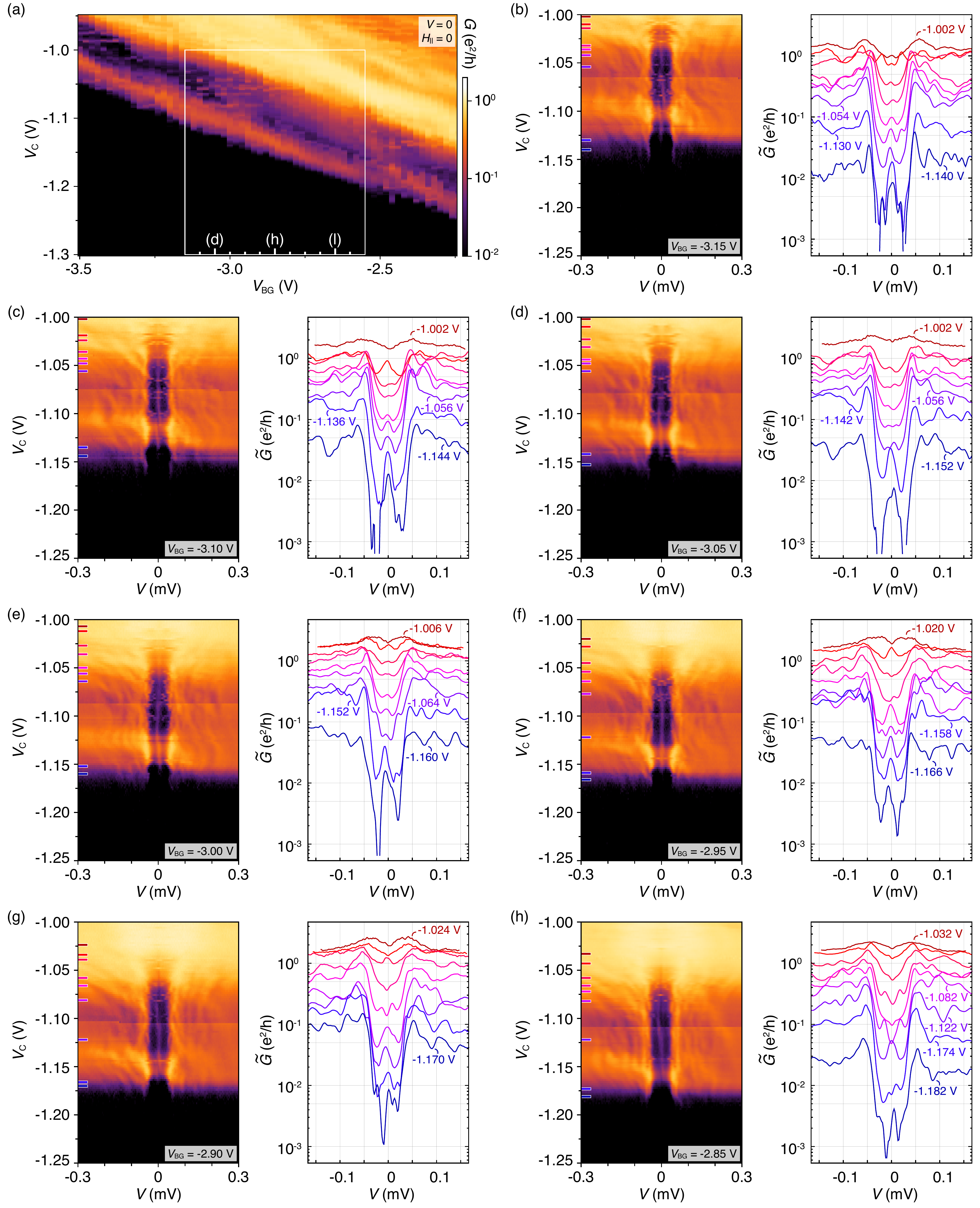}
\end{figure}

\begin{figure}[p!]
\includegraphics[width=\linewidth]{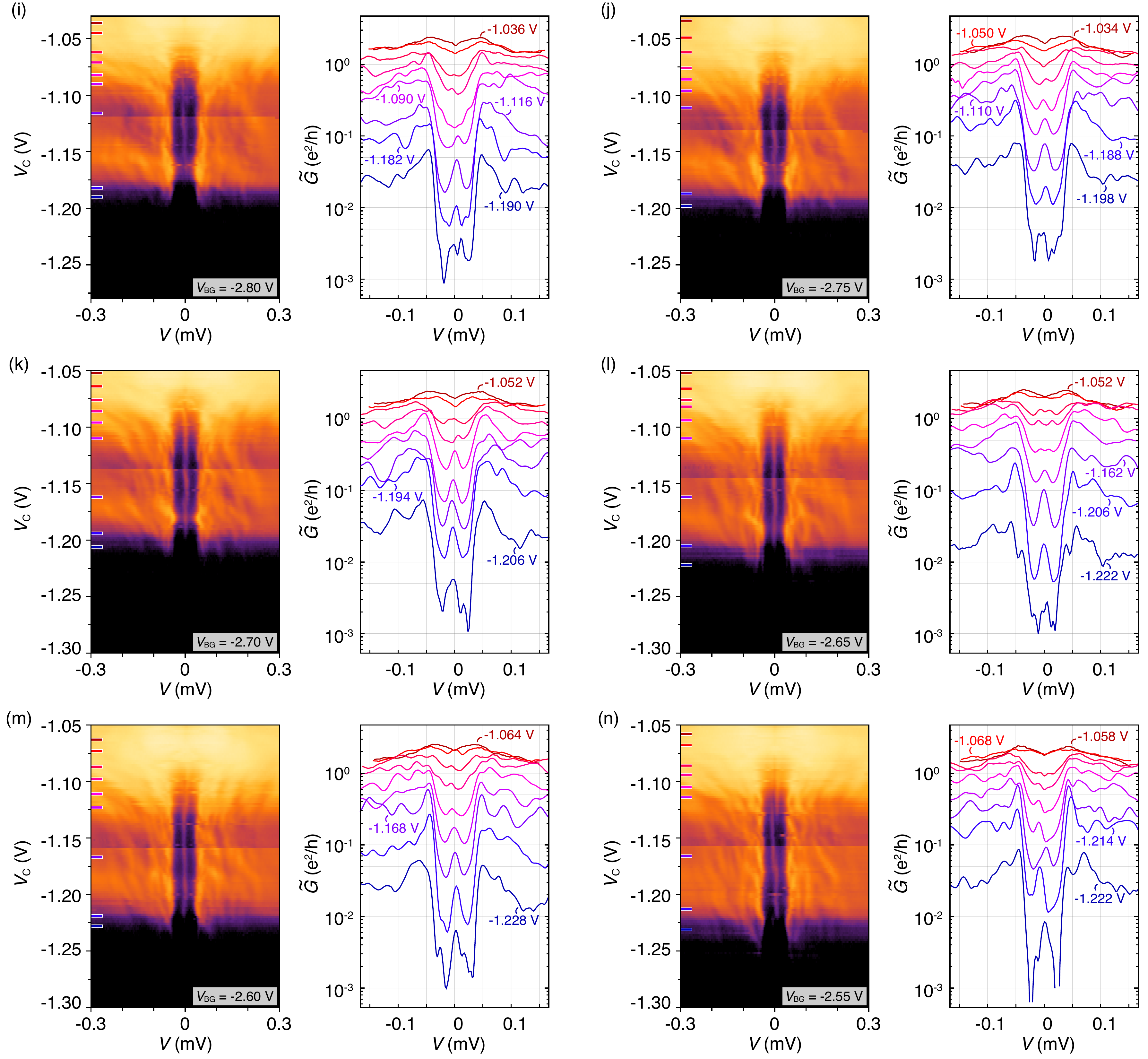}
\caption{\label{fig:S5}
(a) Differential conductance, $G$, measured for device 2 at zero bias as a function of barrier-gate voltage, $V_{\rm C}$, and back-gate voltage, $V_{\rm BG}$.
(b)~Left: $G$ as a function of source-drain bias voltage, $V$, and $V_{\rm C}$ measured at $V_{\rm BG} = -3.15$~V. Right: Line-cut plots of $G$ with subtracted line resistance, $\widetilde{G}$, (see Methods) taken from the right panel at various $V_{\rm C}$ values.
(c--n) Similar to (b) measured within the white box in (a) at various $V_{\rm BG}$ values ranging from $-3.10$~V to $-2.55$~V every $0.05$~V.
All the sweeps display a zero-bias peak that evolves into a zero-bias dip, with the crossover-conductance values varying from $0.2 \, e^2/h$ to above $e^2/h$ depending on $V_{\rm BG}$.}
\end{figure}

\begin{figure}[p!]
\includegraphics[width=\linewidth]{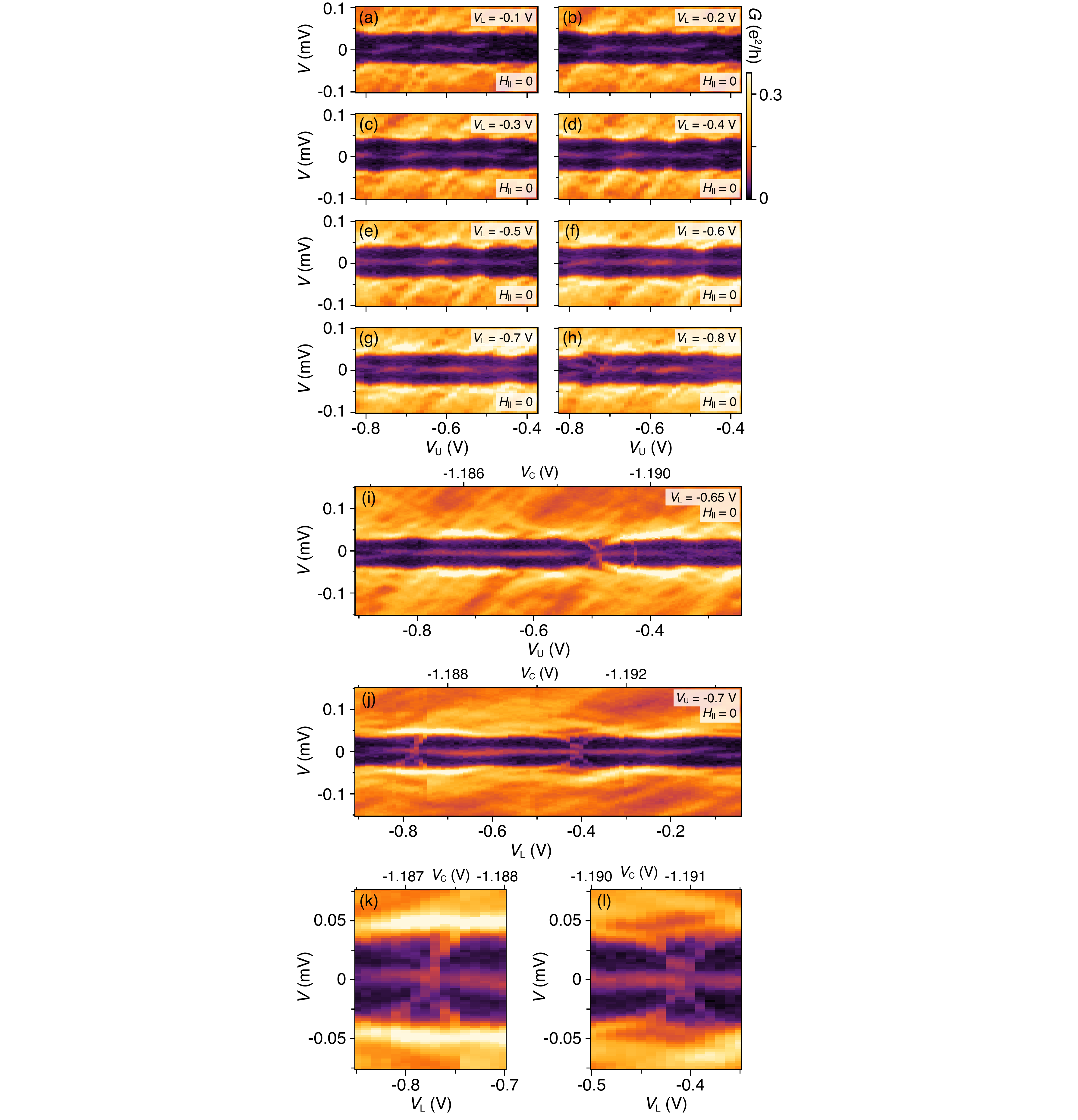}
\caption{\label{fig:S6}
(a--h) Differential conductance, $G$, measured for device~2 as a function of source-drain voltage bias, $V$, and upper-gate voltage, $V_{\rm U}$, taken at various lower-gate voltage, $V_{\rm L}$, values. A pair of faint subgap states is visible for $V_{\rm L}$ close to zero. As $V_{\rm L}$ is reduced, a stable zero-bias peak develops around $V_{\rm L} = -0.6$~V, but then splits again for more negative voltages.
(i) $G$ dependence on $V_{\rm U}$ taken at $V_{\rm L} = -0.65$~V while compensating with $V_{\rm C}$ to maintain barrier iso-potential.
(j) Similar to (i) but as a function of $V_{\rm L}$ taken at $V_{\rm U} = -0.7$~V.
The sharp resonances in (i) and (j) are due to charge motion in the junction and do not split the zero-bias peak.
(k, l) Zoom-ins on the end-state resonances visible in (j).
All the measurements were taken at back-gate voltage $V_{\rm BG} = -3$~V and zero applied magnetic field $H_\parallel = 0$.
}
\end{figure}

\begin{figure}[h!]
\includegraphics[width=\linewidth]{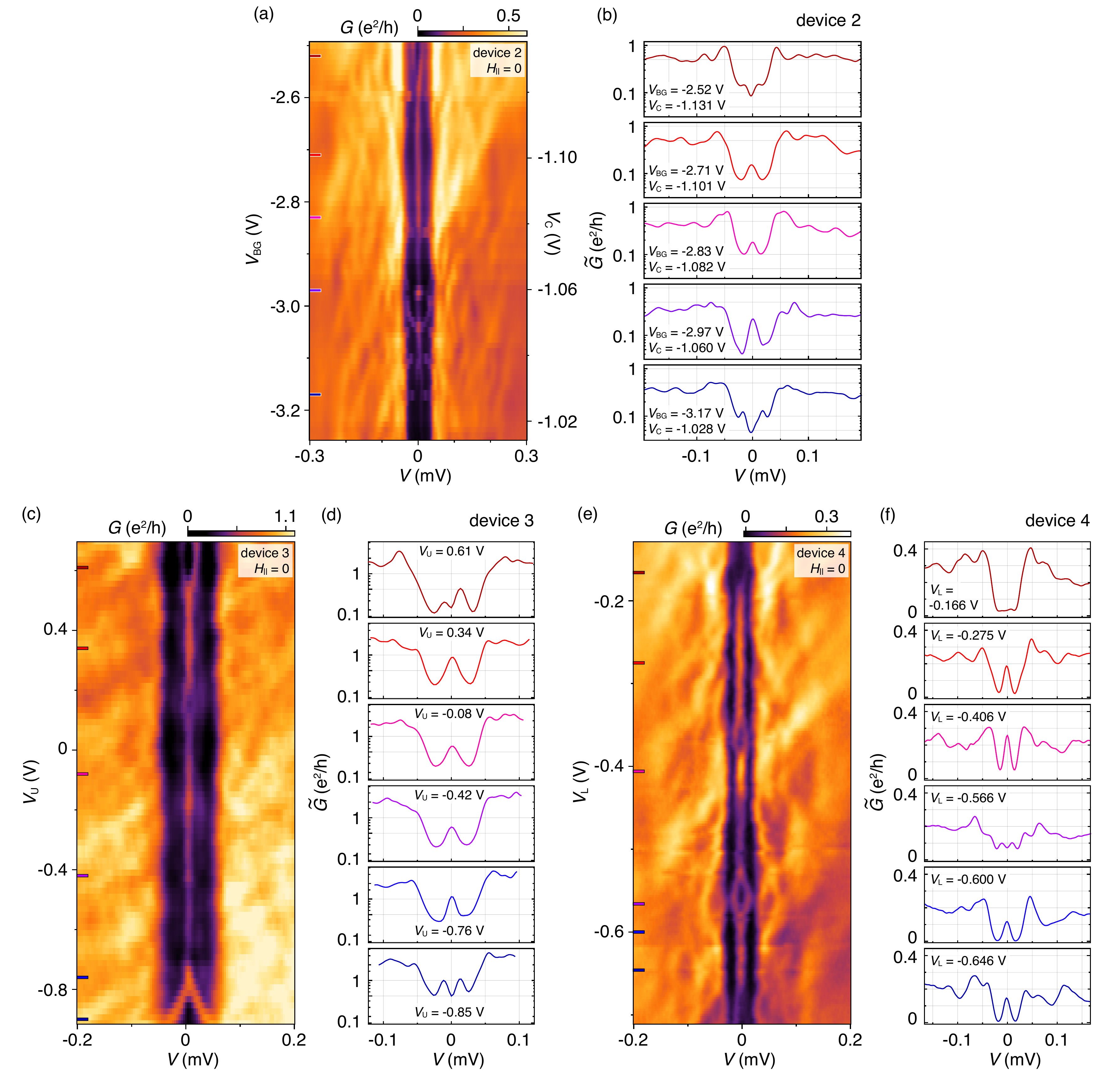}
\caption{\label{fig:S7}
(a) Differential conductance, $G$, measured for device~2 as a function of source-drain bias voltage, $V$ and back-gate voltage, $V_{\rm BG}$, measured along white dashed line in Fig.~\ref{fig:2}(b).
(b) Line-cut plots of $G$ with subtracted line resistance, $\widetilde{G}$, (see Methods) taken from (a) at various $V_{\rm BG}$ values.
(c,d) Similar to (a,b) but for device~3 as a function of upper-gate voltage, $V_{\rm U}$.
(e,f) Similar to (a,b) but for device~4 as a function of lower-gate voltage, $V_{\rm L}$.
All the measurements were taken at zero applied magnetic field $H_\parallel = 0$.
}
\end{figure}

\begin{figure}[p!]
\includegraphics[width=\linewidth]{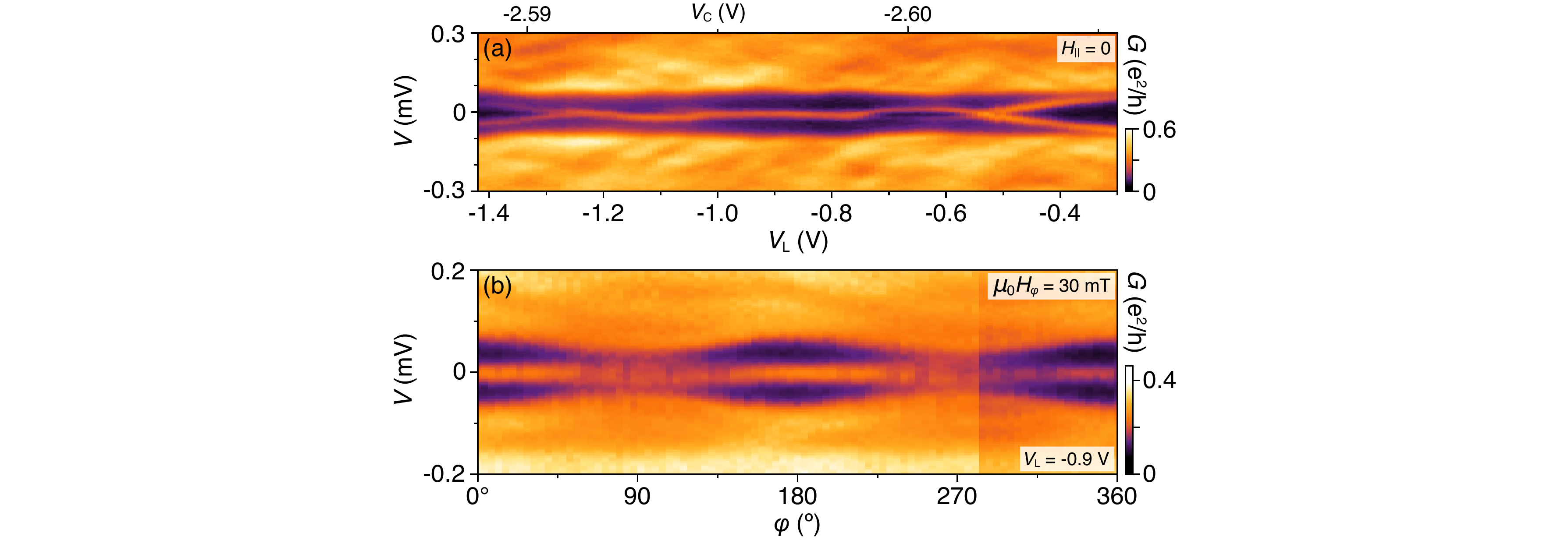}
\caption{\label{fig:S8}
(a) Differential conductance, $G$, measured for device~6 [lithographically equivalent to device 5 shown in Fig.~\ref{fig:2}(a) but with wire length of 1~$\mu$m] as a function of source-drain bias voltage, $V$ and lower-gate voltage, $V_{\rm L}$. Top axis shows compensation gate voltages. The data were taken at zero applied magnetic field $H_\parallel = 0$.
(b) $G$ dependence on in-plane angle, $\varphi$ [see Fig.~\ref{fig:2}(a) for orientation], taken at a fix external magnetic field amplitude, $\mu_{0} H_\varphi = 30$~mT and a gate configuration corresponding to $V_{\rm L} = -0.9$~V in (a) shows a zero-bias peak robust for all $\varphi$.  
}
\end{figure}

\begin{figure}[p!]
\includegraphics[width=\linewidth]{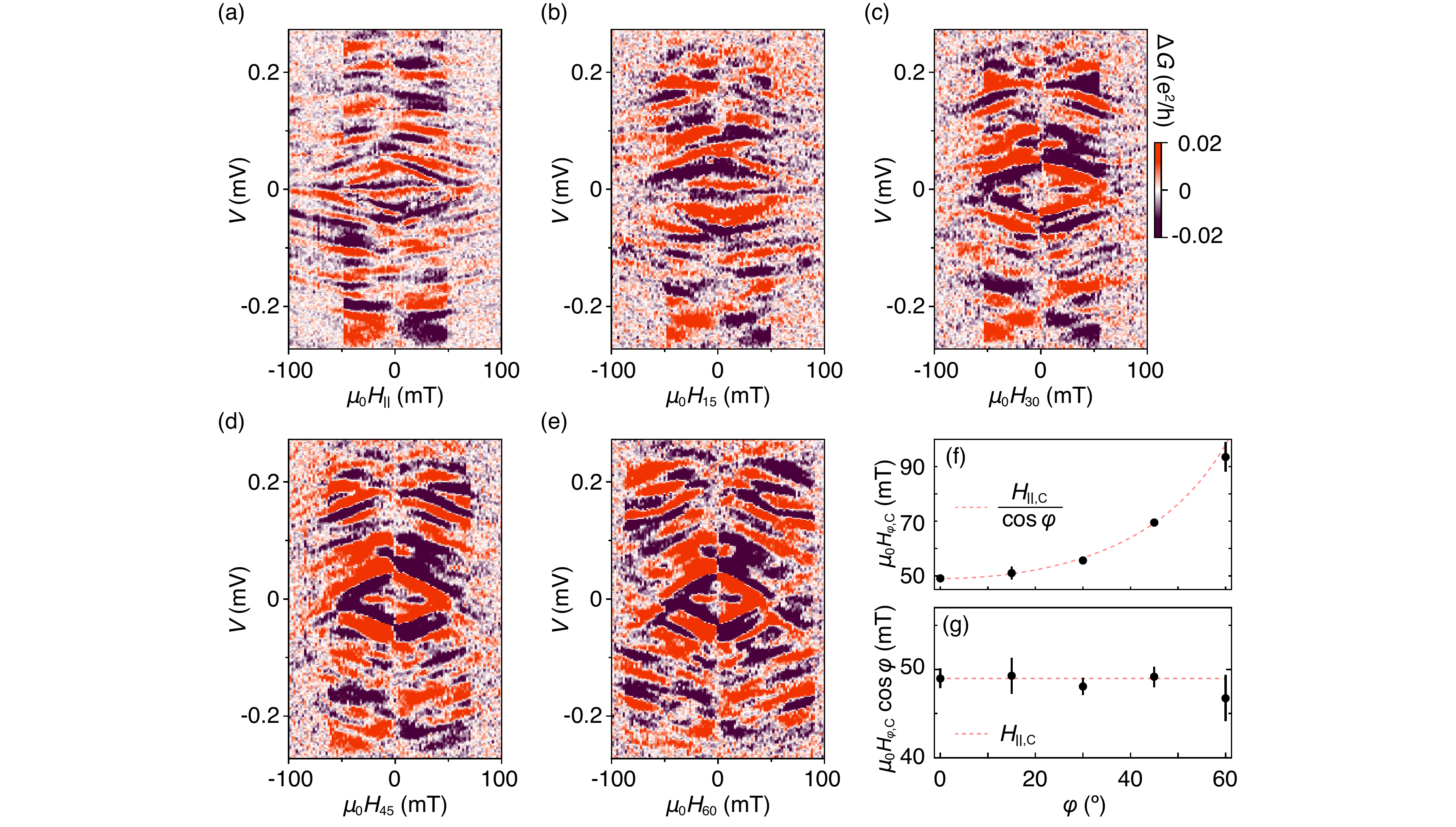}
\caption{\label{fig:S9}
(a) Difference of two conductance maps measured for device 2 with axial magnetic field, $H_\parallel$, swept to opposite directions, as a function of source-drain bias voltage, $V$, showing coercive field $H_{\rm \parallel, C} = \pm 49$~mT.
(b--e) Similar to (a) but with field, $H_\varphi$, swept at in-plane angle, $\varphi$ [see Fig.~\ref{fig:2}(a) for orientation]. The amplitude of coercive field, $H_{\rm \varphi, C}$, increases with $\varphi$.
(f) $H_{\rm \varphi, C}$ measured from (a--e) as a function of $\varphi$ increases as $1/\cos({\varphi})$.
(g) Same data as in (f) multiplied by $\cos ({\varphi})$ does not depend on $\varphi$, indicating that the EuS magnetization is along the wire axis and that only the $H_\parallel$ component controls magnetization.
}
\end{figure}

\begin{figure}[p!]
\includegraphics[width=\linewidth]{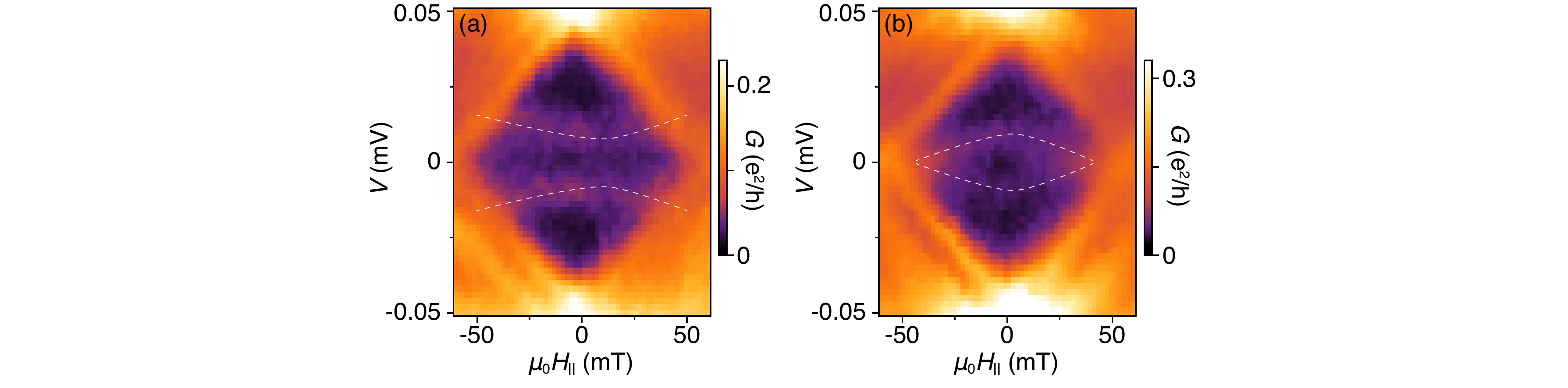}
\caption{\label{fig:S10}
(a) Differential conductance, $G$, as a function of source-drain bias voltage, $V$ and axial magnetic field, $H_\parallel$ for device~2 measured at $V_{\rm L} = -0.75$~V and $V_{\rm R} = -0.6$~V.
(b) Similar to (a) but measured at $V_{\rm L} = -0.65$~V and $V_{\rm R} = -1.23$~V. The applied axial field can act to either increase or decrease the splitting of subgap states. 
}
\end{figure}
\end{document}